\documentclass[11pt]{article}
\usepackage{moriond,epsfig}
\usepackage{amssymb}

\def\Journal#1#2#3#4{{#1} {\bf #2}, #3 (#4)}

\def\PLB{{\em Phys. Lett.}  B}

\def\PRD{{\em Phys. Rev.} D}

\def\be{\begin{equation}}
\def\ee{\end{equation}}
\def\bea{\begin{eqnarray}}
\def\eea{\end{eqnarray}}

\begin{document}
\title{Quantum Chromo (many-body) Dynamics probed in the hard sector at RHIC}

\author{DAVID D'ENTERRIA}

\address{Nevis Laboratories, Columbia University\\ 
Irvington, NY 10533, and New York, NY 10027, USA}

\maketitle\abstracts{
The most significant experimental results on hard processes in
heavy-ion collisions at RHIC collider energies are reviewed. 
Emphasis is put on measurements that provide insights on 
strongly interacting media like the ``Quark Gluon Plasma'' 
and the ``Color Glass Condensate''.
}

\section{Introduction}

Nucleus-nucleus collisions at relativistic energies aim at the 
study of the fundamental theory of the strong interaction, 
Quantum Chromo Dynamics (QCD), at extreme energy densities. 
Two aspects are of particular interest in this physics program:
\begin{enumerate}
\item The production and study under laboratory conditions of the 
plasma of quarks and gluons (QGP), a deconfined and chirally
symmetric state of strongly interacting matter, predicted by QCD 
calculations on the lattice for values of the energy 
density $\epsilon\approx$ 1 GeV/fm$^3$~\cite{latt}.
\item The study of the (non-linear) evolution of the gluon density 
at small values of (Bjorken) fractional momentum $x$ in the
nuclear \footnote{The parton density in a high-energy (Lorentz-contracted) 
nucleus is enhanced by a $A^{1/3}$ factor compared to that in the proton 
and, thus, one has an experimental access to the ($x,Q^2$) kinematical regime 
where higher-twist ($g+g$ fusion) processes are expected to set in 
and saturate the rapidly rising gluon distribution function at small-$x$.} 
(and, in general, hadronic) wave functions,
as described e.g. in the ``Color Glass Condensate'' (CGC) framework~\cite{cgc}.
\end{enumerate}

In hadronic collisions, the production of particles with high transverse 
momentum (jets, single hadrons with $p_{T} \gtrsim$ 2 GeV/$c$, prompt $\gamma$) 
or large mass (heavy quarks) 
results from hard parton-parton scatterings 
with large momentum transfer $Q^2$ (``hard processes''). 
Such production processes provide, thus, direct information 
on the fundamental (quark and gluon) degrees of freedom of QCD.
Since hard cross-sections can be theoretically computed by perturbative 
methods via the collinear factorization theorem~\cite{factor},
inclusive high $p_T$ hadroproduction, jets, 
direct photons, and heavy flavors, have long been considered 
sensitive and well calibrated probes of the small-distance 
QCD phenomena. This paper reviews the most interesting 
results on hard processes from $Au+Au$ reactions at RHIC collider 
energies ($\sqrt{s_{_{NN}}}$ = 200 GeV), and compares them to more 
elementary reactions either in the ``vacuum'' ($p+p$, $e^+e^-$) or 
in a cold nuclear matter environment ($d,l+A$). Several substantial
differences are found which are indicative of important initial- and final- state 
effects in $Au+Au$ reactions that can be directly connected to the properties 
of the QCD medium in which the hard scattering process has taken place.

\section{Hard processes in the QCD vacuum: $p+p$ collisions \small{@} $\sqrt{s}$ = 200 GeV}

\begin{figure}[htbp]
\begin{center}
\vspace{-0.4cm}
\epsfig{file=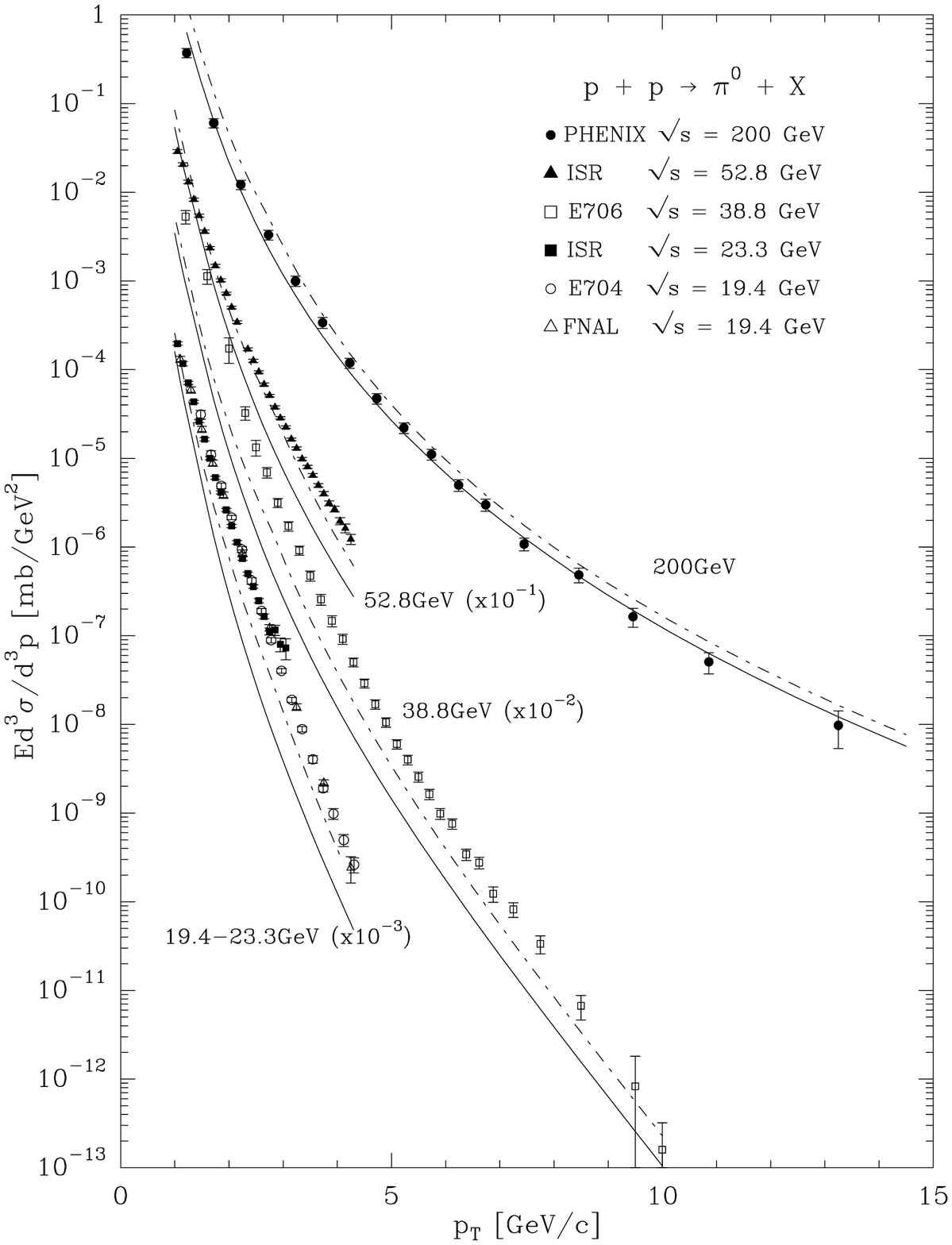,width=7cm,height=8cm}
\epsfig{file=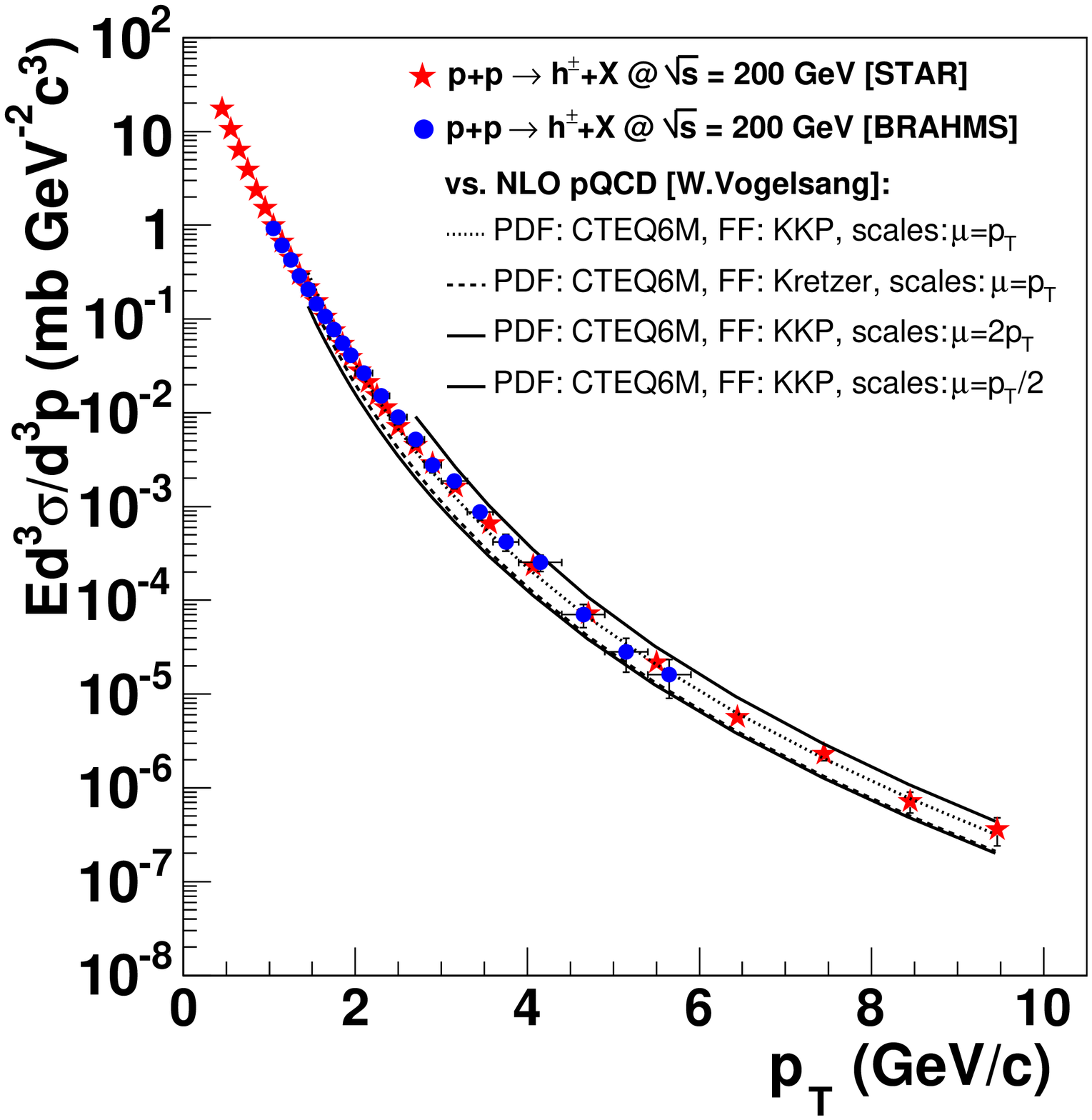,height=7.6cm,width=6.8cm}
\end{center}
\vspace{-0.3cm}
\caption{Invariant cross-sections at midrapidity as a function of $p_T$ 
in $p+p$ collisions at $\sqrt{s}$ = 200 GeV
compared to NLO pQCD calculations (with different scales $\mu = p_T/2, p_T, 2p_T$
and fragmentation functions) for: $p+p\rightarrow\pi^0+X$ (PHENIX
data~\protect\cite{phnx_pp_pi0_200} compared to results at other
$\sqrt{s}$~\protect\cite{aurenche_bourre}, left), and $p+p\rightarrow h^\pm+X$ 
(STAR~\protect\cite{star_pp_chhad_200} and BRAHMS~\protect\cite{brahms}, right).}
\label{fig1}
\end{figure}

Hard processes in proton-proton collisions provide the baseline 
``free space'' reference to which one compares heavy-ion results in order
to extract information about the properties of color many-body dynamics. 
At RHIC, the differential cross-sections in $p+p$ collisions at $\sqrt{s}$ = 200 GeV
for neutral pions~\cite{phnx_pp_pi0_200} and charged hadrons~\cite{star_pp_chhad_200,brahms} 
above $p_T$ = 2 GeV/$c$ are well reproduced by standard 
next-to-leading-order (NLO) pQCD calculations (Figure~\ref{fig1}).
This is at variance with measurements at lower center-of-mass energies
($\sqrt{s}\approx$ 20 -- 63 GeV at CERN-ISR and $\sqrt{s}\approx$ 20 -- 40 GeV
in fixed-target at FNAL, Fig.~\ref{fig1} left), where the $p_T<$ 5 GeV/$c$ 
cross-sections are underpredicted~\cite{aurenche_bourre} 
by pQCD calculations (even supplemented with soft-gluon resummation 
corrections~\cite{resumm}), and additional non-perturbative effects 
(e.g. intrinsic $k_T$~\cite{e706_kt}) must be introduced to bring parton 
model analysis into agreement with data. Hard production in $p+p$ collisions 
at RHIC collider energies seems to be free of the 
``distortive'' non-perturbative effects that are important at lower 
energies, and constitutes thus a experimentally and theoretically well 
calibrated baseline for heavy-ion studies.

\vspace{-0.4cm}
\section{Hard processes in dense QCD matter: 
central $Au+Au$ reactions$\,\mbox{\small{@}}\,\sqrt{s_{\mbox{\tiny{\it{NN}}}}}$=200 GeV}

\subsection{QCD factorization for hard cross-sections in $A+A$ collisions.}

In the absence of initial and final state interactions, 
QCD factorization (based on the implicit premise of incoherent parton 
scattering\footnote{Incoming quarks and gluons undergoing hard scattering 
are ``free'' in a collinear factorized approach, i.e. 
{\it with regard to hard production} the density of partons in a nucleus 
with atomic number $A$ should be equivalent to the superposition of $A$ 
independent nucleons, or $f_{a/A}(x,Q^2) = A\, f_{a/p}(x,Q^2)$ in terms of 
parton distribution functions.}) implies that inclusive A+B cross-sections for 
hard processes should scale simply as $A\cdot B$ times the corresponding $p+p$ 
cross-sections:
\begin{equation}
E\,d\sigma_{AB\rightarrow h}^{hard}/d^3p=A\cdot B \cdot E\,d\sigma_{pp\rightarrow h}^{hard}/d^3p .
\label{eq:AB_scaling}
\end{equation}
More generally, for a given centrality bin (or impact parameter $b$) in a nucleus-nucleus 
reaction, the relation between hard cross-sections in $p+p$ and $A+A$ collisions is:
\begin{equation}
E\,dN_{AB\rightarrow h}^{hard}/d^3p\,(b)=\langle T_{AB}(b)\rangle\cdot E\,d\sigma_{pp\rightarrow h}^{hard}/d^3p ,
\label{eq:TAB_scaling}
\end{equation}
where $T_{AB}(b)$ is the Glauber (eikonal) nuclear overlap function 
at $b$. Since the number of nucleon-nucleon ($NN$) collisions at $b$ is: 
$N_{coll}(b) = T_{AB}(b)\cdot \sigma_{pp}^{inel}$, one can write also Eq.~(\ref{eq:TAB_scaling}) 
in terms of invariant yields as: 
$E\,dN_{AB\rightarrow h}^{hard}/d^3p\,(b)=\langle N_{coll}(b)\rangle\cdot E\,dN_{pp\rightarrow h}^{hard}/d^3p$
(``$N_{coll}$ scaling''). Usually, the standard method to quantify the effects 
of the medium in the production of a given hard probe is provided by the 
{\it nuclear modification factor}:
\begin{equation} 
R_{AB}(p_{T},y;b)\,=\frac{\mbox{\small{``hot QCD medium''}}}{\mbox{\small{``QCD vacuum''}}}\,=\,\frac{d^2N_{AB}/dy dp_{T}}{\langle T_{AB}(b)\rangle\,\times\, d^2 \sigma_{pp}/dy dp_{T}},
\label{eq:R_AA}
\end{equation}
which measures the deviation of $A+B$ at 
$b$ from an incoherent superposition of $NN$ collisions.\\

\begin{figure}[htbp]
\begin{center}
\hspace{-1.cm}
\epsfig{file=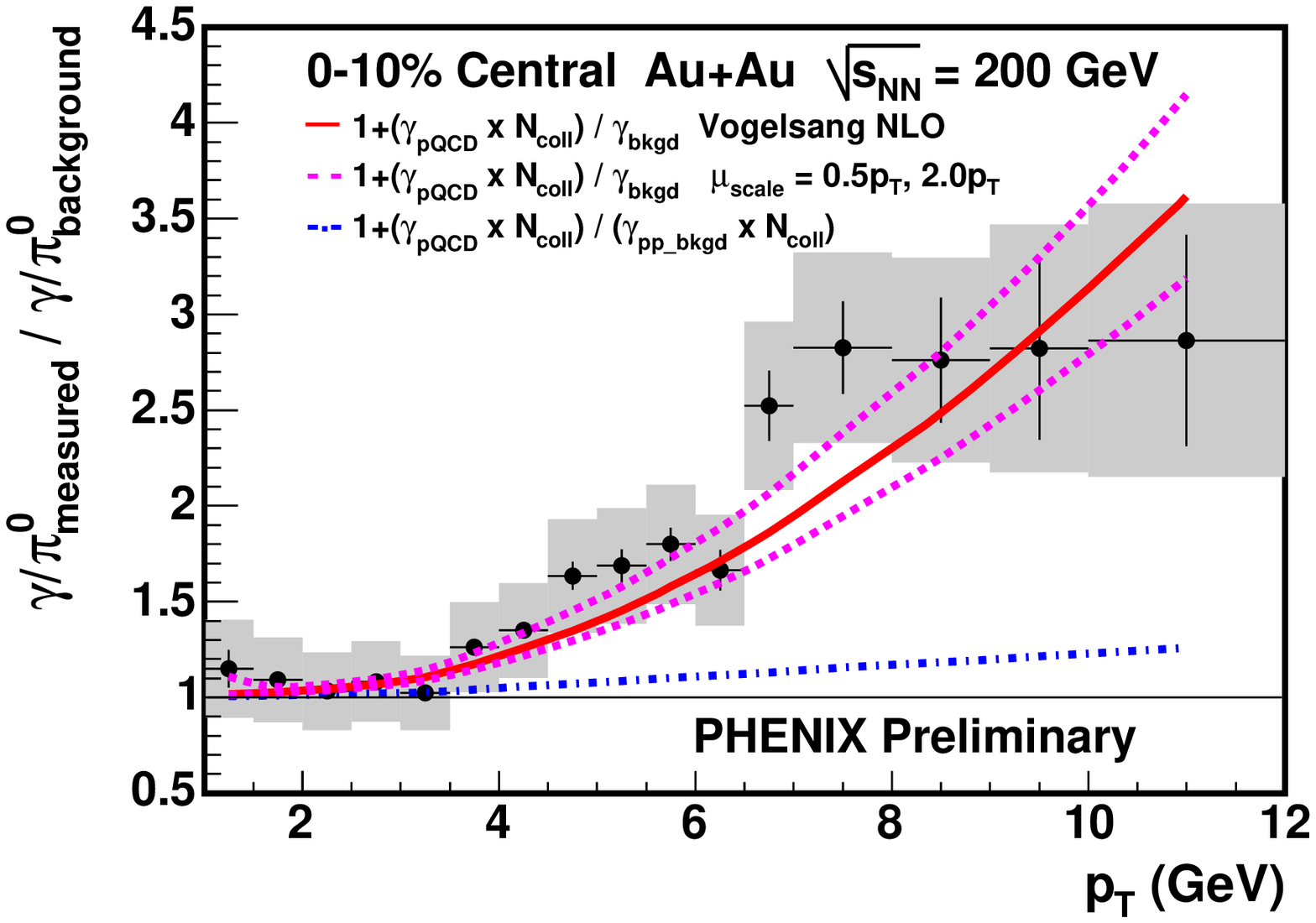,height=6.3cm}
\epsfig{file=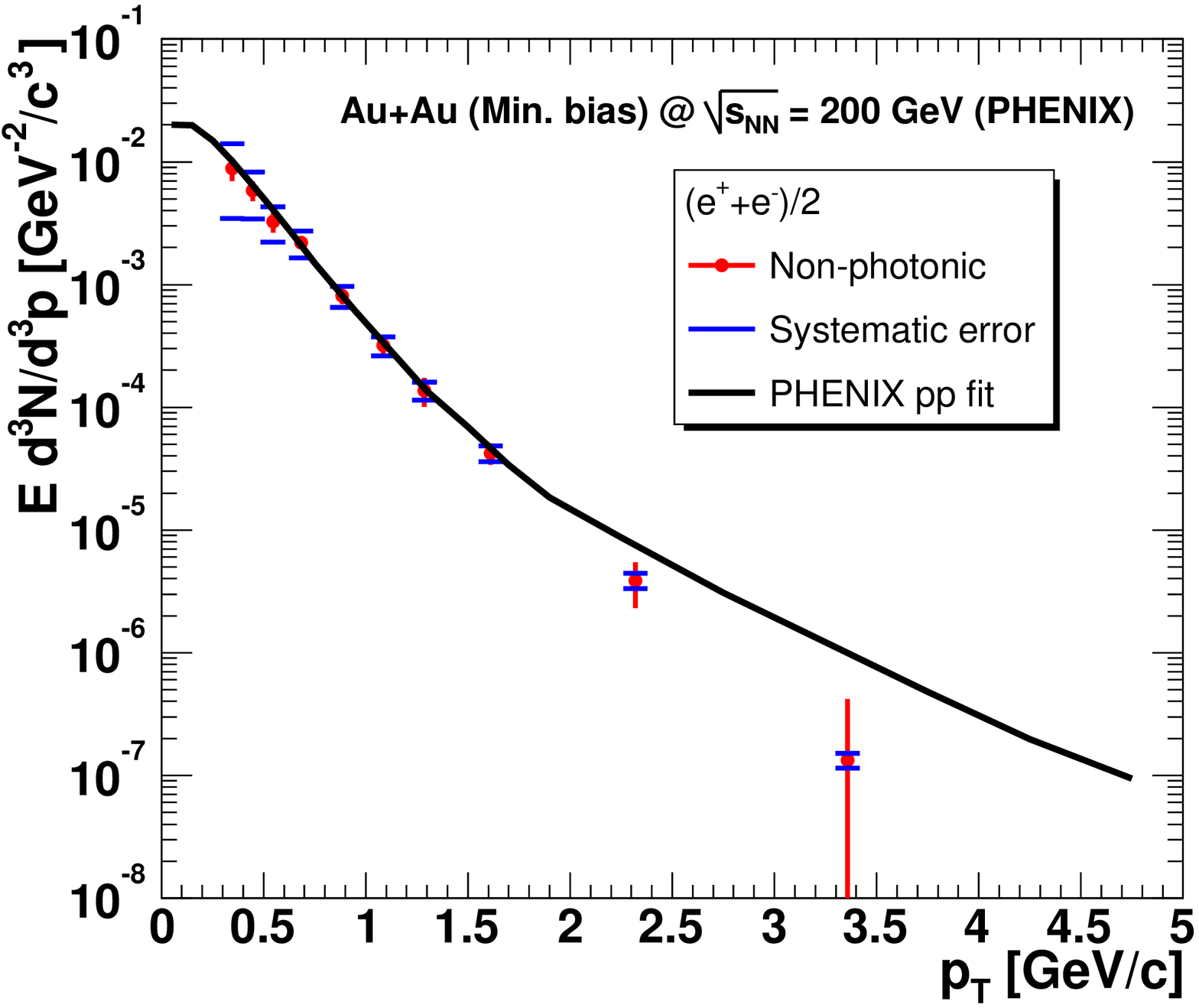,height=6.cm}
\vspace{-0.2cm}
\caption{Hard spectra in $Au+Au$ at $\sqrt{s_{\mbox{\tiny{\it{NN}}}}}$ = 200 GeV:
(a) Direct photon ``excess'' (given by the ratio $dN_{\gamma\,total}/dp_T$ over 
$dN_{\gamma\, decay}/dp_T$, normalized by the $\pi^0$ spectra) measured in central $Au+Au$ 
compared to NLO pQCD yields scaled by $N_{coll}$ (red curve)~\protect\cite{justin}, 
(b) ``Non-photonic'' single electron spectrum (mostly from open charm and beauty semi-leptonic
decays) measured in min.bias $Au+Au$ compared to the $p+p$ spectrum scaled by 
$A^2$ (black curve)~\protect\cite{sean}.}
\label{fig:AuAu_dirgamma_and_elec}
\end{center}
\end{figure}

The expectations of QCD factorization, Eqs. (\ref{eq:AB_scaling})--(\ref{eq:TAB_scaling}), 
are indeed fulfilled by hard processes in $Au+Au$ 
collisions at RHIC. The measured spectra of direct photons and single electrons 
from heavy quark decays (Figure~\ref{fig:AuAu_dirgamma_and_elec}) 
is consistent with the $N_{coll}$ (or $A^2$) scaling expected for
hard scattering in the absence of medium effects$^c$.

\vspace{-0.2cm}
\subsection{High $p_T$ suppression of (light flavored) hadron spectra in central $Au+Au$.}
\begin{figure}[htbp]
\begin{center}
\epsfig{file=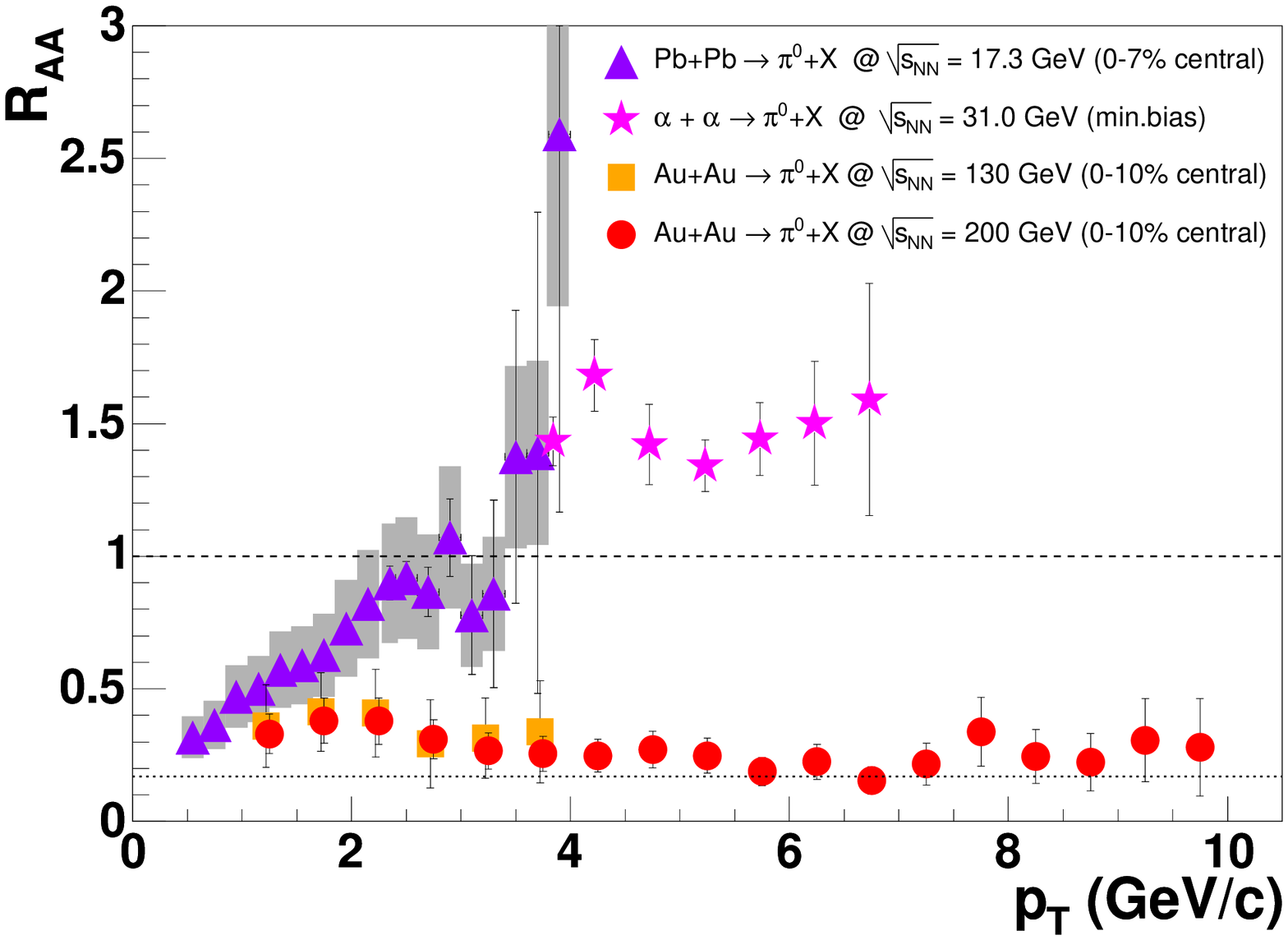,height=5.1cm}
\epsfig{file=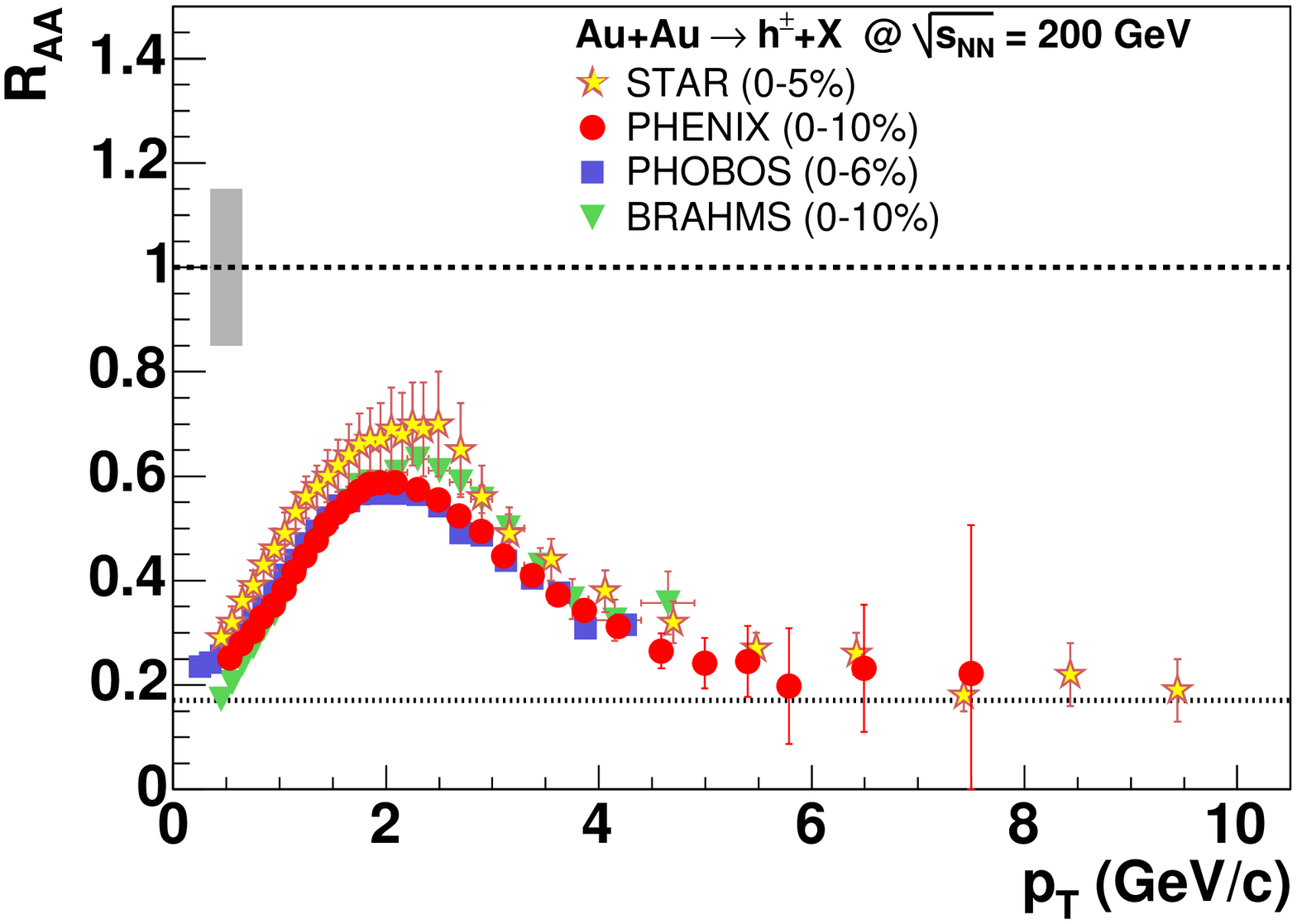,height=5.2cm}
\end{center}
\caption{Nuclear modification factors $R_{AA}(p_T)$, Eq. (\protect\ref{eq:R_AA}), for: 
(i) $\pi^0$ measured in central nucleus-nucleus reactions at 
SPS~\protect\cite{wa98,denterria}, ISR~\protect\cite{ISR_pi0}
and RHIC~\protect\cite{ppg003_ppg014} (left);
(ii) inclusive charged hadrons in central $Au+Au$ at RHIC. 
[The dashed (dotted) lines are the expectation of ``$N_{coll}$ ($N_{part}$) 
scaling'' for hard (soft) particle production.]}
\label{fig:R_AA_pi0_chhad}
\end{figure}

High $p_T$ light flavored hadrons emitted in $Au+Au$ collisions result from 
the fragmentation of hard scattered (light) quarks and gluons and, thus, 
their yields should follow the $N_{coll}$ scaling for incoherent parton 
scattering seen in direct $\gamma$ and heavy-quark cross-sections
\footnote{Note that although the yield of single $e^{\pm}$ 
in $Au+Au$ seems to be suppressed at high $p_T$ with respect to the $p+p$ 
scaled reference (Fig.~\ref{fig:AuAu_dirgamma_and_elec}, right), 
the overall differential and {\it integrated} cross-sections~\cite{sean} 
are consistent with $N_{coll}$ scaling in the whole $p_T$ range covered, 
as expected for $e^\pm$ coming from hard (heavy-quark) processes.}. 
One of the most significant results at RHIC so far is the observation that
the high $p_{T}$ cross-sections of neutral pions and inclusive charged hadrons 
in central $Au+Au$ are strongly suppressed compared to these expectations.
Figure~\ref{fig:R_AA_pi0_chhad} (left) shows $R_{AA}$ as a function of $p_T$ for 
$\pi^0$ produced in nucleus-nucleus reactions at different center-of-mass energies. 
RHIC data at 200 GeV (circles) and 130 GeV (squares)~\cite{ppg003_ppg014} are 
noticeably below unity in contrast to the ``Cronin'' enhanced production observed 
in $\alpha+\alpha$ collisions at CERN-ISR~\cite{ISR_pi0} (stars) and to the
CERN-SPS central Pb+Pb data~\protect\cite{wa98} consistent, within errors, 
with $N_{coll}$ scaling
~\protect\cite{denterria} (triangles).
The same factor of 4--5 suppression is observed above $p_T\approx$ 5 GeV/$c$ in
the spectra of charged hadrons (Fig.~\ref{fig:R_AA_pi0_chhad}, right).
This marked breakdown of the expectations from collinear factorization in 
central $Au+Au$ collisions, together with the unsuppressed production 
of colorless hard probes (direct photons, Fig.~\ref{fig:AuAu_dirgamma_and_elec}, left),
is clearly indicative of strong final-state effects affecting high $p_T$
$Au+Au$ hadroproduction 
at RHIC. In ``jet quenching'' scenarios~\cite{jet_quenching}, 
the scattered partons lose energy via gluon radiation in the dense partonic 
system formed in the reaction and the resulting high $p_T$ (leading) 
hadrons have a reduced energy compared to standard fragmentation in the 
vacuum. Theoretical calculations~\cite{jet_quenching} reproduce the 
observed suppression pattern assuming the formation of a strongly 
interacting (expanding) system with very large initial gluon densities 
$dN^{g}/dy\approx$ 1100.

\subsection{Suppression of back-to-back dijet correlations in central $Au+Au$.}
\begin{figure}[htbp]
\begin{center}
\begin{tabular}{cc}
   \epsfig{file=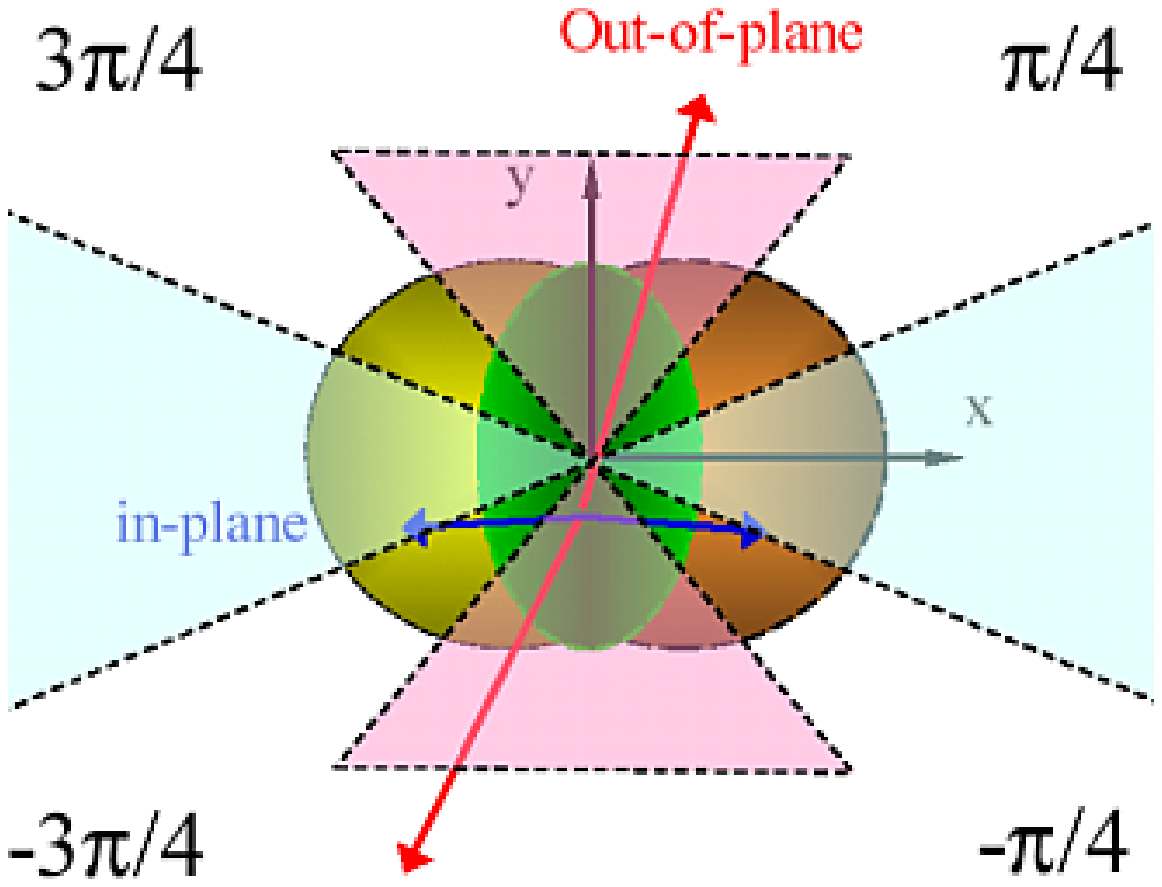,height=5.5cm} & 
   \epsfig{file=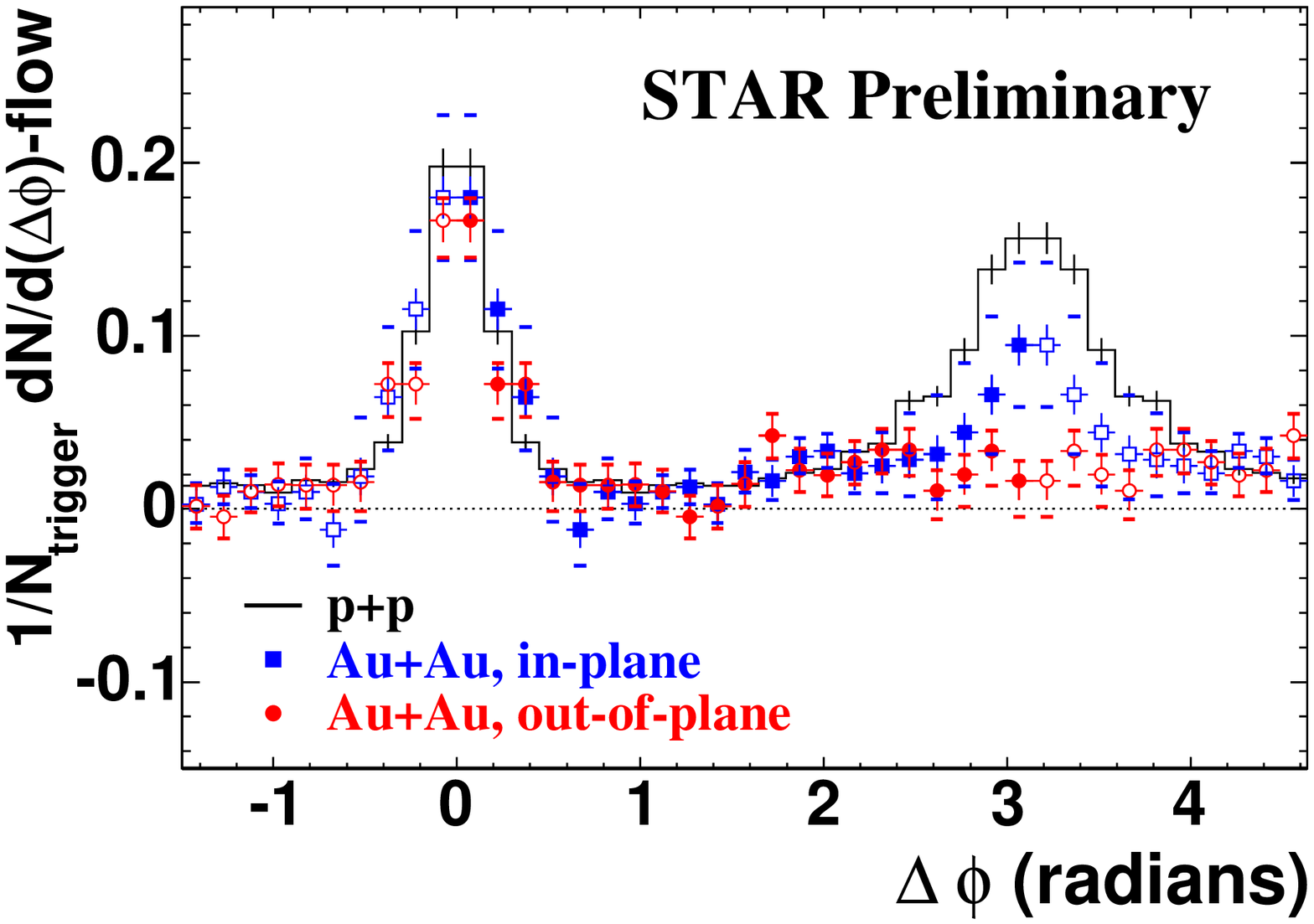,height=5.5cm} \\
\end{tabular}
\end{center}
\vspace{-0.5cm}
\caption{Azimuthal distributions of associated particles for (high $p_T$) 
trigger particles emitted in-plane (squares) and out-of-plane (circles)  
in $Au+Au$ collisions at centrality 20\%-60\%, compared to the $p+p$ 
reference (histo)~\protect\cite{star_awayside}.}
\label{fig:inOut_star}
\end{figure}

A second prominent RHIC result is the disappearance of jet-like azimuthal 
correlations in the away-side hemisphere of trigger particles with 
$p_T$ = 4 -- 6 GeV/$c$~\cite{star_awayside}. 
Figure~\ref{fig:inOut_star} (right) shows the azimuthal distribution of 
associated particles (with 2~GeV$/c < p_T < p_T^{\rm trig}$) in $Au+Au$ 
(after subtraction of the collective elliptic flow component) 
compared with $p+p$ reference data. 
The near-side ($|\Delta\phi|\lesssim$ 0.5) correlations measured in $Au+Au$ 
are clearly jet-like as those in $p+p$ collisions. The back-to-back correlations 
($|\Delta\phi-2\pi|\lesssim$ 0.7) in $Au+Au$ collisions for trigger particles 
emitted parallel to the reaction plane (``in plane'') are suppressed 
compared to $p+p$, and even more suppressed for the ``out of plane'' 
trigger particles. Such behavior can be explained by jet quenching 
models~\cite{jet_quenching}, where the energy loss of a parton 
depends on the distance traveled through the dense medium (note that in 
non-central reactions the overlap nucleus-nucleus region has an almond-like 
shape with shorter length in the in-plane than in the out-of-plane direction, 
Fig.~\ref{fig:inOut_star} left).

\subsection{Enhanced baryon production at intermediate $p_T$.}
\begin{figure}[htbp]
\begin{center}
\epsfig{file=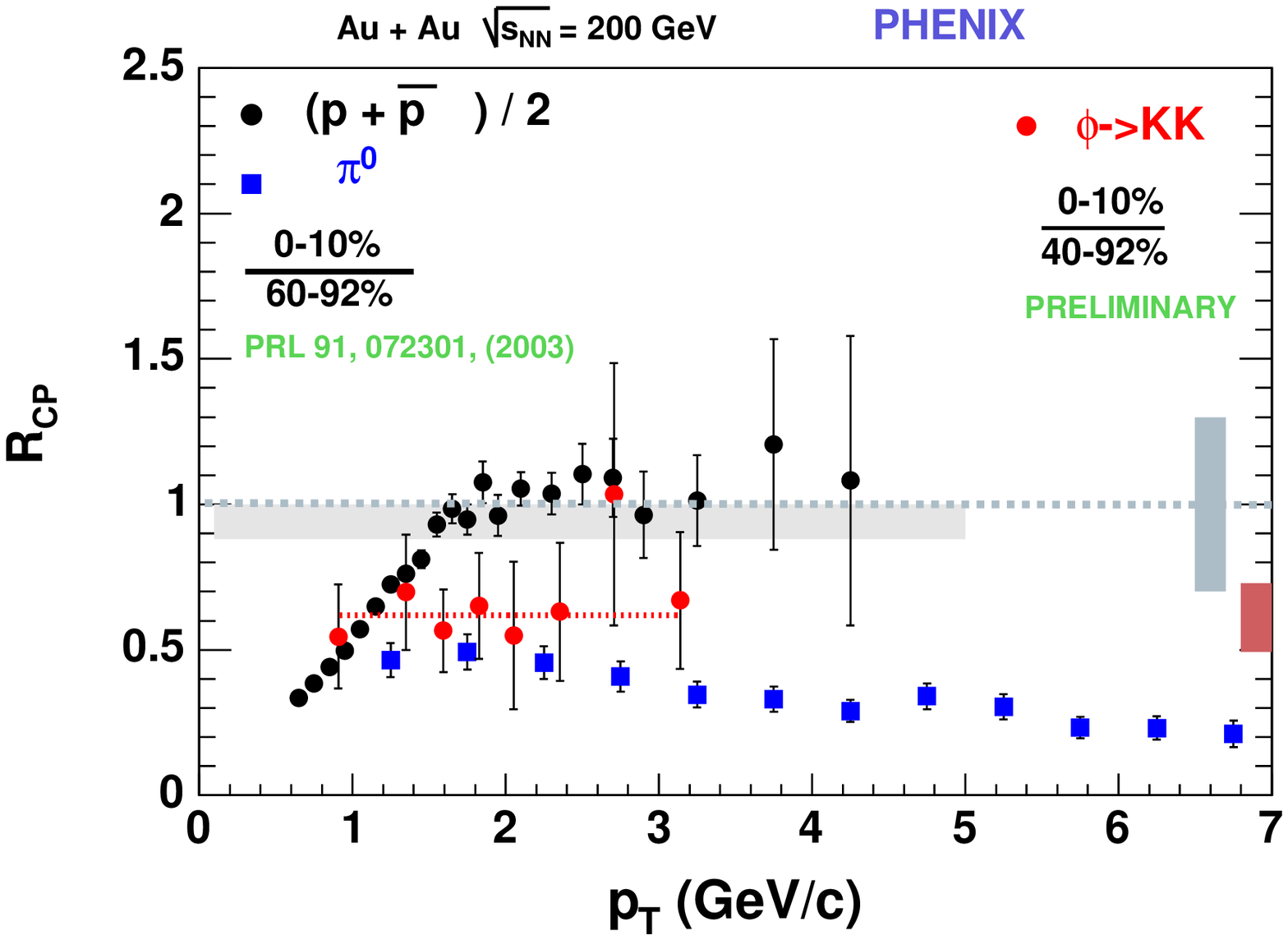,height=5.6cm,width=7.85cm}
\epsfig{file=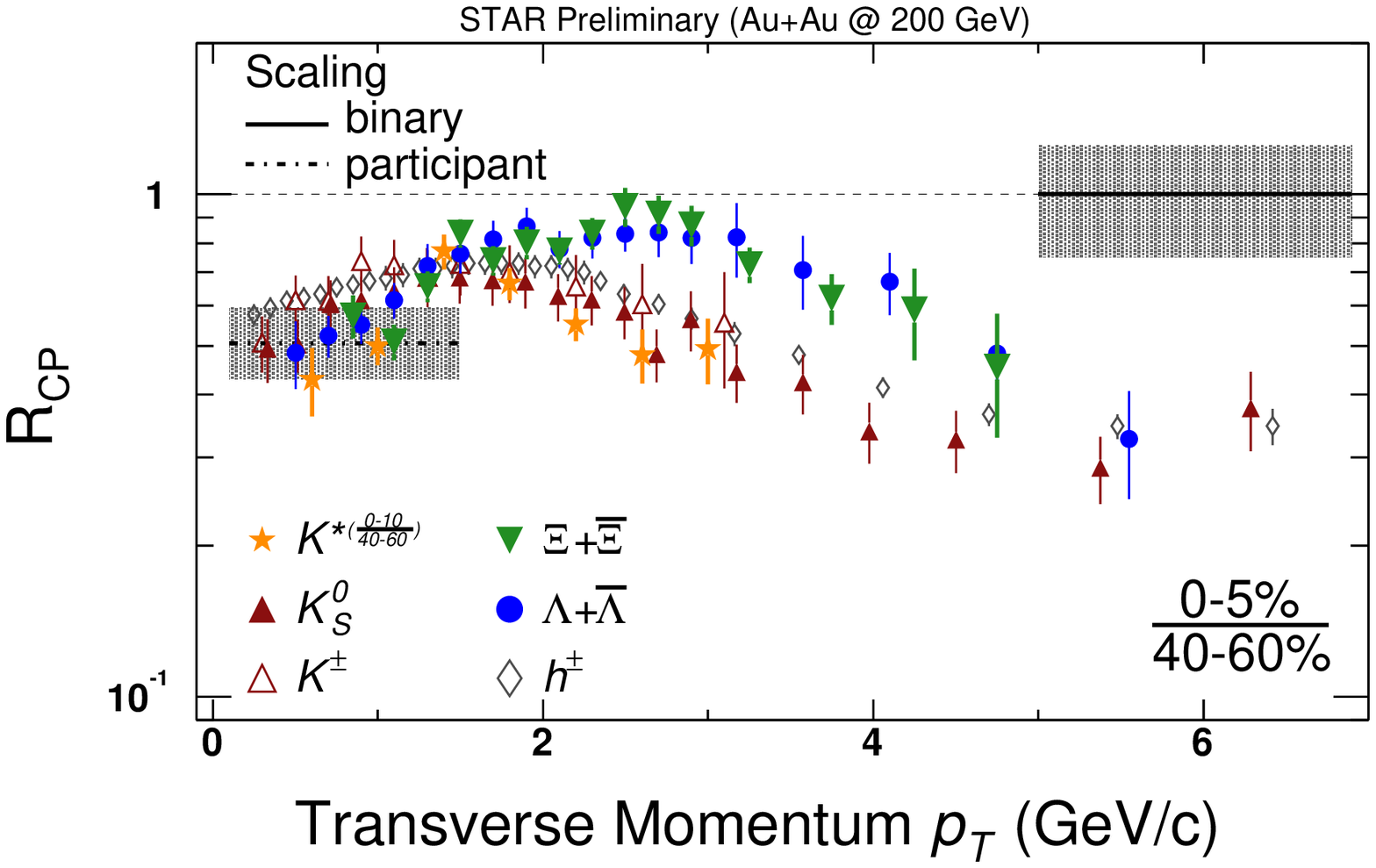,height=5.6cm,width=7.85cm}
\end{center}
\vspace{-0.4cm}
\caption[]{Ratio of central over peripheral $N_{coll}$-scaled yields, $R_{cp}$, 
as a function of $p_{T}$ for different baryons and mesons identified
in $Au+Au$ collisions at 200 GeV by PHENIX~\protect\cite{phenix_baryons} 
(left) and STAR~\protect\cite{star_baryons} (right).}
\label{fig:flavor_dep}
\end{figure}

A third intriguing observation at RHIC is the different suppression 
pattern for baryons and mesons at moderate $p_T$ values.
Fig.~\ref{fig:flavor_dep} shows the $N_{coll}$-scaled central 
to peripheral ratio\footnote{Since the peripheral $Au+Au$ 
(inclusive and identified) spectra scale with $N_{coll}$ when compared 
to the $p+p$ yields, the ratio $R_{cp}$ carries basically the same 
information as the nuclear modification factor $R_{AA}$.}, 
$R_{cp}$, for baryons and mesons measured by PHENIX~\cite{phenix_baryons} 
(left) and STAR~\cite{star_baryons} (right). 
In the range $p_T\approx$ 2 -- 4 GeV/$c$, (anti)baryons 
(p, $\Lambda$, $\Xi$) are not (or barely) suppressed ($R_{cp}\sim$ 1) 
at variance with mesons ($\pi^{o,\pm}$, $K^\pm$, $K^0_s$, $\phi$) 
which are reduced by a factor of 2 -- 3. 
The resulting p/$\pi\sim$1 ($\Lambda/K_0^s\sim$1.5) ratio in this $p_T$ range 
is clearly at odds with the ``perturbative'' p/$\pi\sim$0.2 ($\Lambda/K^0_s\sim$0.5) 
value measured in $p+p$ or $e^{+}e^{-}$ collisions. Such a particle 
composition is inconsistent with standard fragmentation 
functions, and points to an additional non-perturbative mechanism for 
baryon production in central $Au+Au$ in this intermediate $p_T$ range. 
Enhanced production due to mass-dependent boost effects (e.g. due to
collective hydrodynamic flow) seems to be excluded since mesons as heavy 
as the proton ($\phi$, $K^\star$) are as equally suppressed as lighter mesons.
Recombination models~\cite{fries}, on the other hand, where hadronization
occurs mainly via quark coalescence, predict such enhanced baryon over 
meson yields (due simply to the extra momentum gained by the addition of 
a third constituent quark). Beyond $p_{T} \approx 5$ GeV/$c$ 
fragmentation becomes the dominant production mechanism for all species 
in agreement with the data.

\section{Hard spectra in cold QCD matter: $d+Au$ collisions \small{@} $\sqrt{s_{\mbox{\tiny{\it{NN}}}}}$ = 200 GeV}

\subsection{High $p_T$ production at midrapidity: Cronin enhancement}
\label{sect:dAu_midrapidity}
\begin{figure}[htbp]
\begin{center}
\epsfig{file=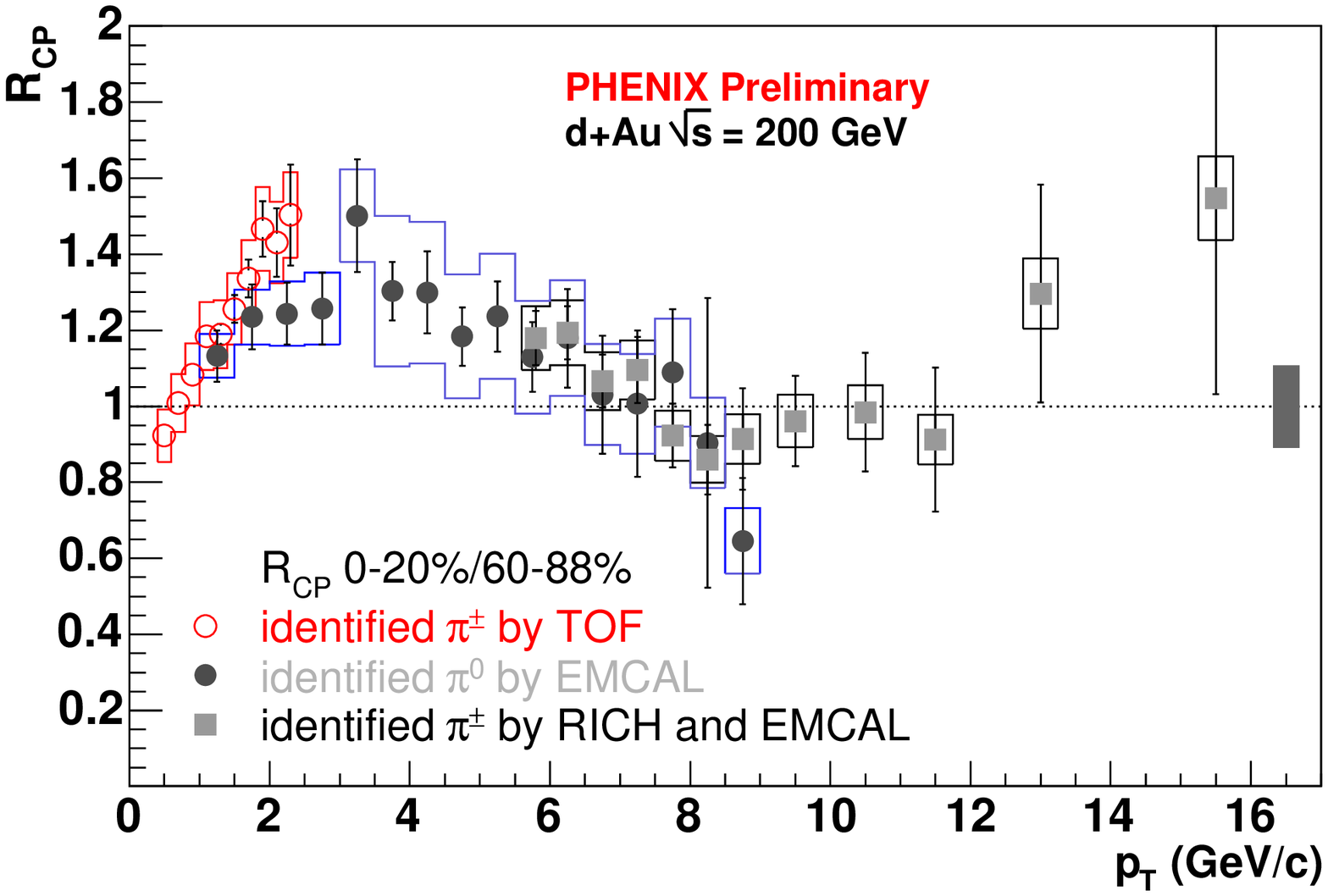,height=5.6cm,width=8.5cm}
\epsfig{file=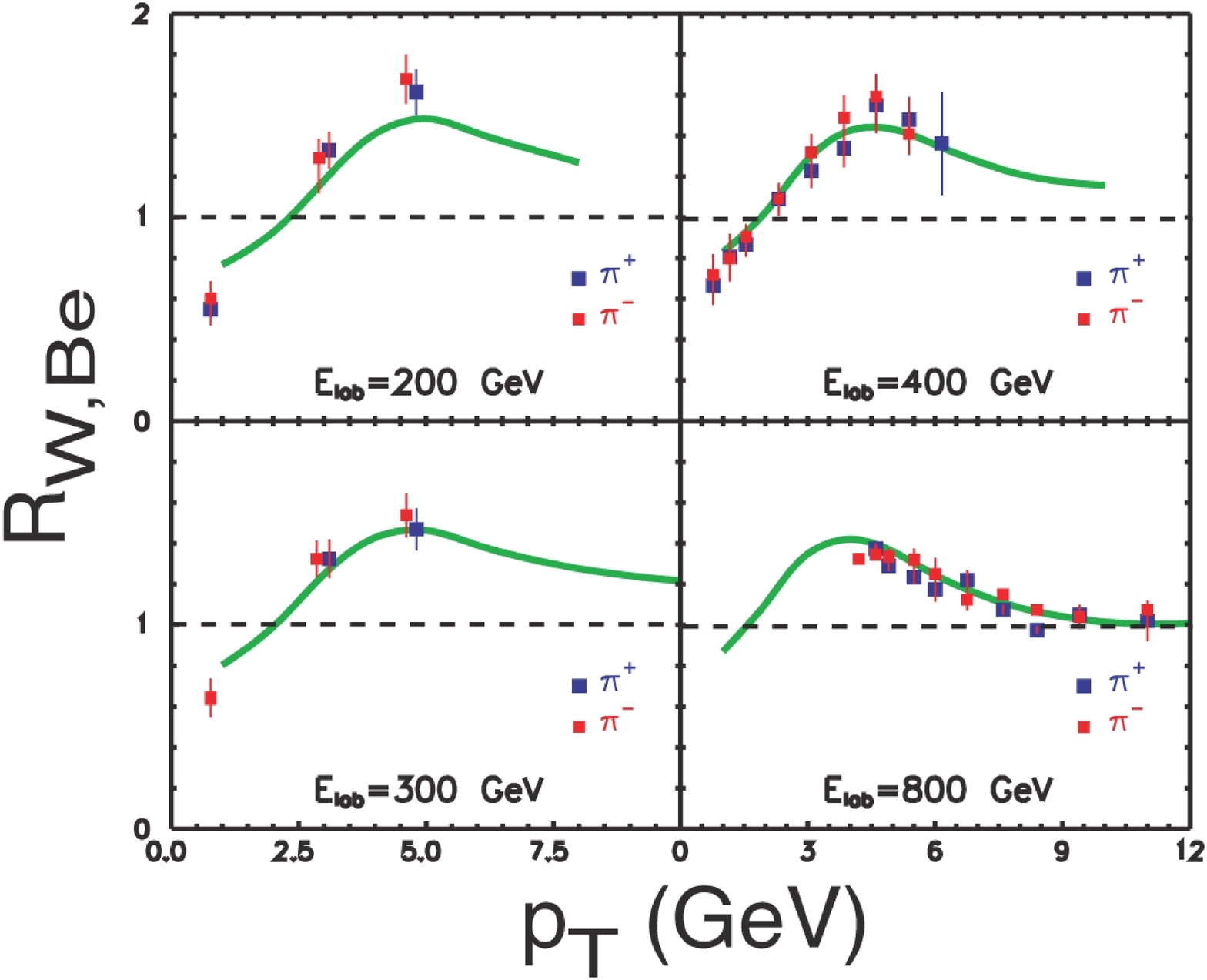,height=5.5cm,width=7.3cm}
\end{center}
\caption{Left: Ratio of central over peripheral $N_{coll}$-scaled 
yields, $R_{cp}$, versus $p_{T}$ for pions measured 
in $d+Au$ at $\sqrt{s_{_{NN}}}$ = 200~GeV. Right: Ratio of pion
yields measured in $p+W$ over $p+Be$ at $\sqrt{s_{_{NN}}}\approx$ 20 -- 40~GeV~\protect\cite{cronin}.
Note that the Cronin effect disappears above $p_T\approx$ 8 GeV/$c$
(the data approaches the perturbative limit $R_{cp}\approx$1).}
\label{fig:cronin_dA_pA}
\end{figure}

High $p_T$ pion production at y = 0 in $d+Au$ collisions 
at $\sqrt{s_{_{NN}}}$ = 200~GeV (Fig.~\ref{fig:cronin_dA_pA}) 
is not suppressed but enhanced ($R_{cp}>$ 1) compared to the expectations 
of QCD factorization. Such a result was first observed in $p+A$ fixed-target 
experiments at $\sqrt{s_{_{NN}}}$ = 20 -- 40~GeV (``Cronin effect''~\cite{cronin})
(Fig.~\ref{fig:cronin_dA_pA}, right) and is usually interpreted in terms of 
multiple soft and semi-hard interactions which broaden the transverse 
momentum of the colliding partons on their way in/out the traversed
nucleus. This result confirms that the high $p_T$ deficit in central
$Au+Au$ collisions (Fig.~\ref{fig:R_AA_pi0_chhad}) is not an initial-state 
effect arising from strong modifications of the gluon distribution function 
in nuclei as proposed by CGC approaches, but results instead from a 
final-state effect in the produced dense medium.

\subsection{High $p_T$ production at forward rapidities: searching for gluon saturation effects}
\begin{figure}[htbp]
\begin{center}
\epsfig{file=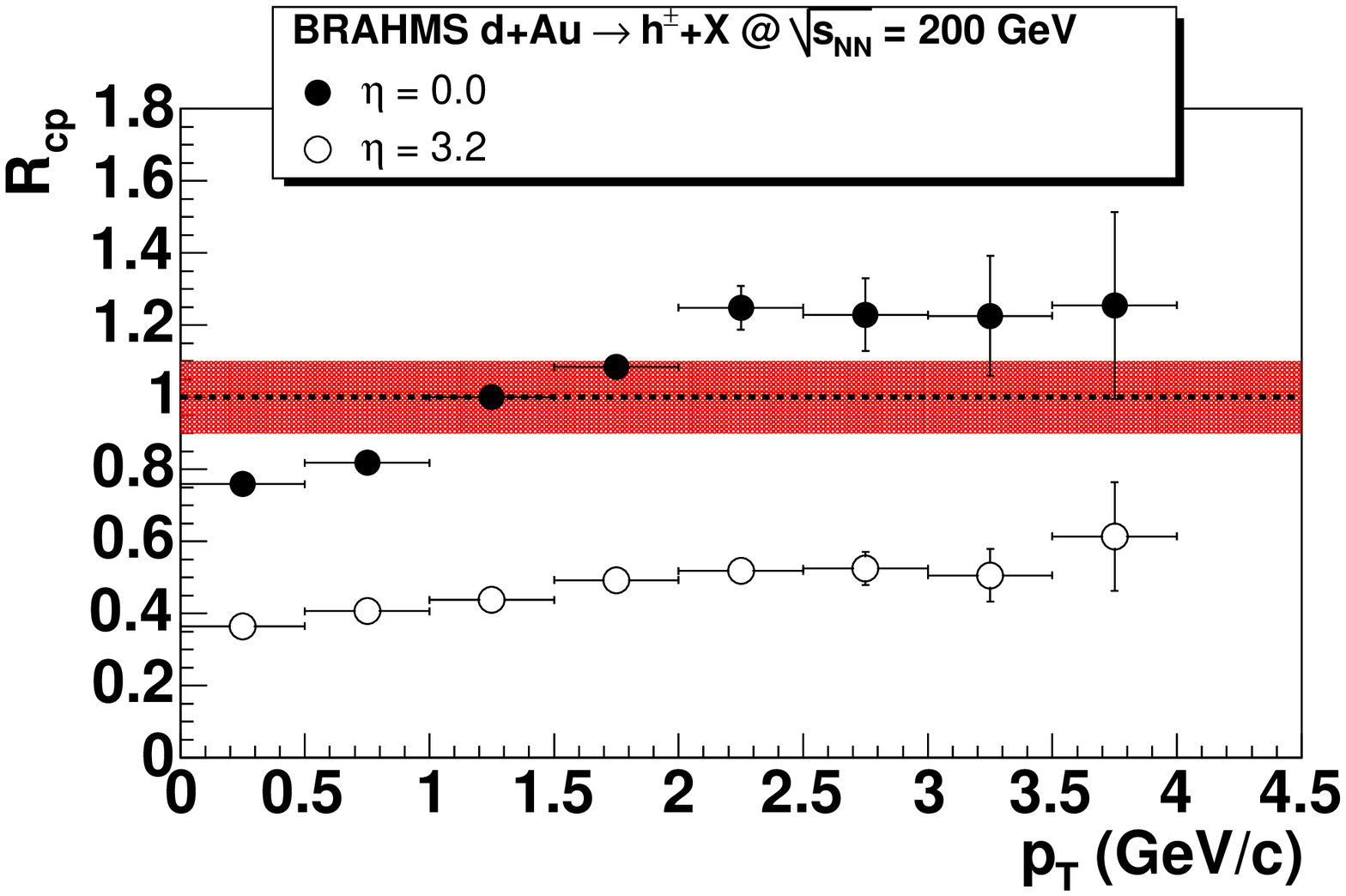,width=7.5cm,height=5.5cm}
\hspace{0.5cm}
\epsfig{file=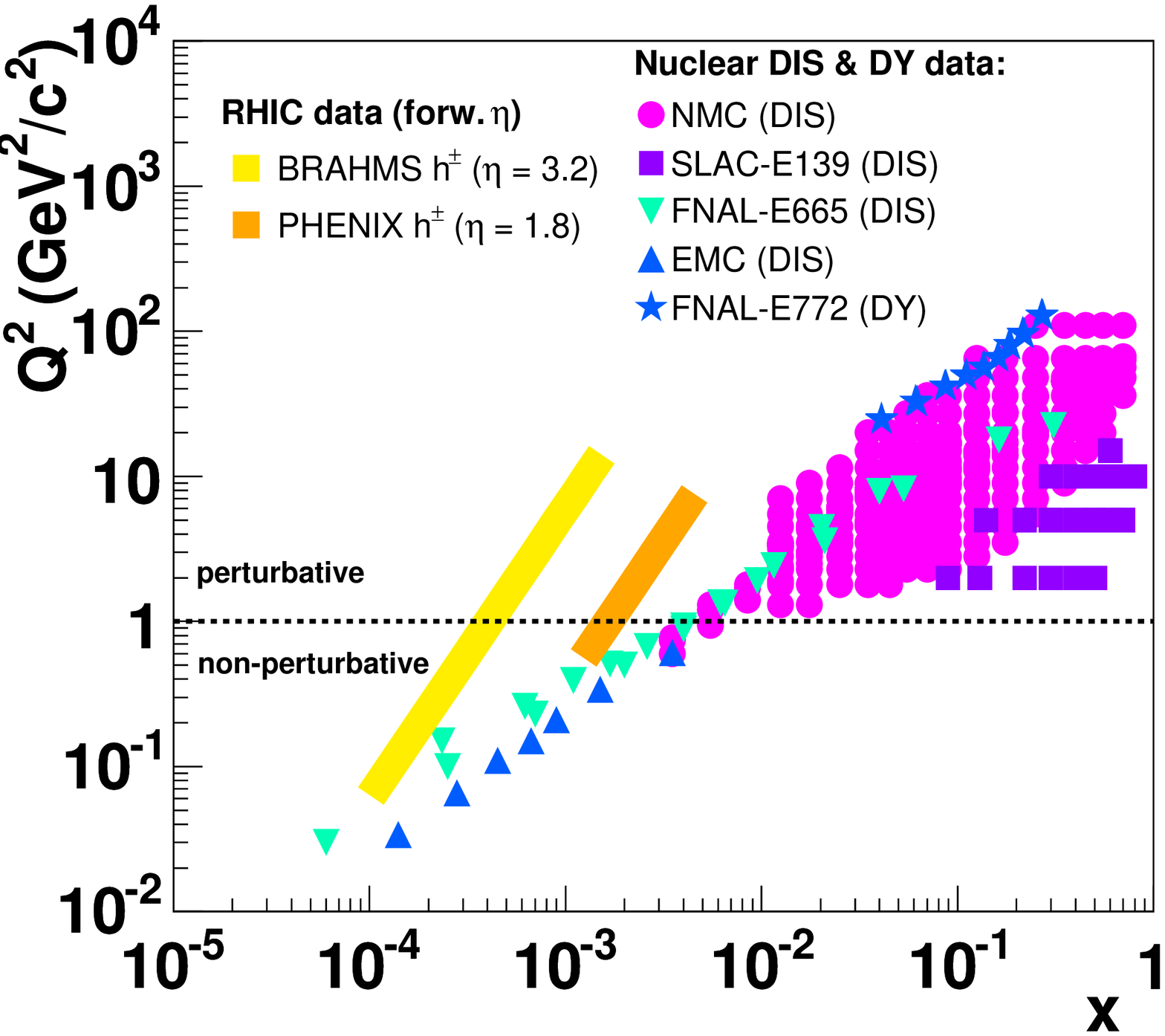 ,width=6.8cm,height=6.0cm}
\end{center}
\caption{Left: Ratio of central over peripheral $N_{coll}$-scaled 
yields, $R_{cp}$, versus $p_{T}$ for charged hadrons measured by 
BRAHMS at $\eta$ = 0 (dots) and $\eta$ = 3.2 (open circles) 
in $d+Au$ at $\sqrt{s_{_{NN}}}$ = 200~GeV~\protect\cite{brahms}. 
Right: Kinematical $x$-$Q^2$ domain probed in nuclear DIS and DY 
at fixed-target energies, and in $d+Au$ at forward rapidities at RHIC.}
\label{fig:Rcp_brahms}
\end{figure}

At variance with $Au+Au$ collisions, lepton- and proton- (or deuteron-) nucleus 
collisions allow to probe the nuclear parton distribution functions (PDF) 
with minimal distortions due to final-state effects. 
Small values of parton fractional momenta, $x_2$, in the $Au$ 
nucleus (where gluons are overly dominant) can be probed studying 
hard production in the forward direction. 
Since $x_{1,2} = p_T/z\sqrt{s}(e^{\pm y_1}+e^{\pm y_2})$ for a $2\rightarrow 2$ 
scattering, $x$ decreases by a factor of $\sim$10 for every 2-units of rapidity 
one moves away from $y$ = 0. BRAMHS~\cite{brahms} and PHENIX~\cite{mliu} results 
on high $p_T$ charged hadron production at pseudorapidities $\eta$ = 3.2 and 
$\eta$ = 1.8, corresponding to $x\approx$ $\mathcal{O}$(10$^{-4}$) 
and $\mathcal{O}$(10$^{-3}$) respectively, show a suppression instead of 
an enhancement as found at $\eta$ = 0 (Fig.~\ref{fig:Rcp_brahms}, left). 
This is the first time that nuclear ``shadowing'' is observed
at such low-$x$ values in the {\it perturbative} domain 
($Q^2\approx p_T^2>$ 1 GeV$^2/c^2$) (Fig.~\ref{fig:Rcp_brahms}
right). BRAHMS $R_{cp}\approx$ 0.5 result would indicate
that the ratio of $Au$ over $p$ gluon densities is
$R_{G}^{Au}(x\approx 10^{-4},Q^2\approx$ 2 GeV$^2/c^2)\approx$ 0.5, 
whereas standard {\it leading-twist} DGLAP analysis of the 
nuclear PDFs~\cite{nPDF} (based on global fits of 
the DIS and Drell-Yan data above $Q^2$ = 1 GeV$^2/c^2$ shown in 
Fig.~\ref{fig:Rcp_brahms}, right) indicate a less significant 
amount of gluon shadowing in this kinematical range: 
$R_{G}^{Au}\approx$ 
0.8.

\section{Summary}
During its first four years of operation, RHIC has provided 
many new and exciting results on the 
many-body dynamics of QCD at high energies. The suppressed high $p_T$ 
hadroproduction observed in central $Au+Au$ reactions and 
in $d+Au$ collisions at forward-rapidities is inconsistent 
with the basic QCD factorization expectations that describe 
particle production in $p+p$ at $\sqrt{s}$ = 200~GeV. 
The factor of 4--5 suppression in central $Au+Au$ is unambiguously 
due to final-state effects (since no such an effect is seen
in $d+Au$ collisions at $y$ = 0 and electromagnetic probes, 
insensitive to the colored final-state, are not depleted in
$Au+Au$) and can be reproduced by calculations of parton energy 
loss in a strongly interacting medium with energy densities well 
above those where lattice QCD predicts a transition to a Quark 
Gluon Plasma. The factor of $\sim$2 deficit observed at $y\approx$ 3 
in $d+Au$ reactions may be the first empirical indication of higher-twist 
(non-linear) QCD effects at small Bjorken-$x$ in the hadronic
wave functions, as described in ``Color Glass Condensate'' 
approaches.


\end{document}